\documentclass[12pt,letterpaper]{article}

\usepackage{amsmath,amssymb,calc}
\usepackage{graphicx}

\newcommand\be{\begin{equation}}
\newcommand\ee{\end{equation}}
\newcommand{\bea}{\begin{eqnarray}}
\newcommand{\eea}{\end{eqnarray}}

\newcommand{\la}{\langle}
\newcommand{\ra}{\rangle}

\newcommand{\nn}{\nonumber}
\newcommand{\pd}{\partial}

\def\id{\protect{{1 \kern-.28em {\rm l}}}}

\def\unit{\relax{\rm 1\kern-.26em I}}

\def\id{\protect{{1 \kern-.28em {\rm l}}}}

\begin{document}

\begin{titlepage}
\begin{center}
\hfill QMUL-PH-07-18 \\
\vskip 15mm

{\Large {\bf O'KKLT at Finite Temperature \\[3mm] }}

\vskip 10mm

{\bf Lilia Anguelova, Vincenzo Cal\`o}

\vskip 4mm
{\em Center for Research in String Theory}\\
{\em Department of Physics, Queen Mary, University of London}\\
{\em Mile End Road, London, E1 4NS, UK.}\\
{\tt l.anguelova,v.calo@qmul.ac.uk}\\

\vskip 6mm

\end{center}

\vskip .1in

\begin{center} {\bf Abstract }\end{center}

\begin{quotation}\noindent
We study whether finite temperature corrections decompactify the internal space in KKLT compactifications with an uplifting sector given by a system that exhibits metastable dynamical supersymmetry breaking. More precisely, we calculate the one-loop temperature corrections to the effective potential of the volume modulus in the KKLT model coupled to the quantum corrected O'Raifeartaigh model. We prove that for the original KKLT model, namely with one exponent in the non-perturbative superpotential, the finite temperature potential is runaway when at zero temperature there is a dS minimum. On the other hand, for a non-perturbative superpotential of the race-track type with two exponents, we demonstrate that the temperature-dependent part of the effective potential can have local minima at finite field vevs. However, rather unexpectedly, it turns out that these minima do not affect the structure of the full effective potential and so the volume modulus is stabilized at the local minimum of the zero temperature potential for the whole range of validity of the supergravity approximation.
\end{quotation}
\vfill

\end{titlepage}

\eject

\tableofcontents

\section{Introduction}

Moduli stabilization in string compactifications is a long-standing problem whose
resolution started taking shape only rather recently. It was realized in \cite{DRS,GKP}
that an essential ingredient is turning on background fluxes. This leads to the
stabilization of all geometric moduli in type IIA on CY(3) \cite{IIA} and of the complex
structure moduli in type IIB \cite{GKP}. In the latter case, one can stabilize the
remaining K\"{a}hler moduli by taking into account non-perturbative effects \cite{KKLT}
or a combination of perturbative and non-perturbative corrections \cite{Quevedo}. Thus it might seem that type IIA compactifications
are under better control. However, this is not the case because of the backreaction of
the fluxes on the geometry, resulting in the deformation of the initial CY to a
non-K\"{a}hler manifold whose moduli space is not well-understood. On the other hand, in
type IIB there is a class of solutions in which the only consequence of the presence of
nonvanishing fluxes is the appearance of a warp factor (see \cite{Grana} for a
comprehensive review on flux compactifications). This makes the type IIB set-up much more
tractable.

Naturally then, the KKLT proposal \cite{KKLT} for dS vacua in type IIB with stabilized moduli
attracted a lot of attention.\footnote{dS vacua can be obtained in a more straightforward
way in other string compactifications with background flux, as shown for example for the strongly coupled
heterotic string in \cite{BCK} (see \cite{AV} for the $R^4$-corrected version).  However, generically the relevant geometry is much more complicated than a CY with a warp factor.} Unfortunately,
the need to add by hand anti-D3 branes in order to lift the original AdS vacuum to dS
makes it problematic to describe this set-up in supergravity. An improvement that does not
require anti-D3 branes was proposed in \cite{BKQ}. There, the uplifting is achieved by
having nonzero D-terms from world-volume fluxes on D7 branes that wrap a three-cycle in the CY
3-fold. However, because of the relationship between D- and F- terms in supergravity, this
scenario turned out to be difficult to realize \cite{FDcor}; although, see \cite{Dterms} for some recent progress. The above difficulties can be circumvented by coupling the KKLT sector to an ISS
sector or, more generally, to a field theory sector that exhibits dynamical supersymmetry breaking
to a metastable state (for brevity, MDSB) \cite{DPP}; see also \cite{AHKO}.\footnote{Since the work of
\cite{ISS}, many more examples were found in the literature \cite{msDSB}, thus showing
that the phenomenon of MDSB is quite generic in supersymmetric field theories.} In this
way, one has a natural uplifting that is also completely under control in the effective 4d
$N=1$ supergravity description.\footnote{An earlier proposal for F-term uplifting was considered in \cite{LNR}.}

In fact, one can capture the essential features of F-term uplifting due to MDSB in the KKLT set-up by taking the uplifting sector to be the O'Raifeartaigh model. The reason is that many theories with dynamical supersymmetry breaking can be approximated near the origin of field space by this model \cite{CLP}.\footnote{More precisely, this is true for the theories that realize DSB via the mechanism of \cite{OrDSB}.} The O'Raifeartaigh-uplifted KKLT, termed O'KKLT, model was first proposed
and studied in \cite{KL}. It was pointed out there that the original KKLT proposal, i.e.
with one exponent in the superpotential, leads to tension between low scale supersymmetry
breaking and the standard high scale cosmological inflation. This undesirable situation
can be resolved by considering a racetrack-type superpotential with two exponents
\cite{KLBP,KL}. We will study here thermal corrections to the effective potential of the
O'KKLT model with one or with two exponentials.

Temperature corrections to a model with MDSB (namely, the ISS model) were considered in
\cite{ACJK,FKKMT}. The motivation there was to address the question how natural is it for the system
to be in a local minimum with broken susy, given that it has global supersymmetric minima.
It turned out that the metastable minimum is thermodynamically preferable. More precisely,
\cite{FKKMT} showed that if one starts from a local minimum of the effective potential at finite
temperature and considers what happens as the temperature decreases, then one finds that
the system rolls towards the metastable, and not towards a global, vacuum of the
zero-temperature potential. This picture persists upon coupling the ISS model to
supergravity \cite{ART}. Unfortunately though, including the volume modulus $\rho$, i.e.
considering the full KKLT-ISS model, is rather complicated technically. The main reason
is that the K\"{a}hler potential for $\rho$ is not canonical, unlike the K\a"{a}hler
potential for the ISS fields. This leads to rather untractable expressions for the
effective potential at nonzero temperature.\footnote{Initial steps in analyzing the latter were
discussed in \cite{ART}. Also, see \cite{BHLR} for a study at finite $T$ of KKLT with D-term or anti-${\rm D}3$ uplifting in the approximation of treating the moduli as background fields, which do not contribute to thermal loops.} However, one can make a lot of progress by considering the O'KKLT model instead. The latter is
tractable enough to enable us to study the phase structure of the finite temperature
effective potential for the field $\rho$. At the same time, as already mentioned above, it captures all the main features
of the MDSB-uplifted KKLT scenario; see \cite{KL}.\footnote{Clearly, for nonzero temperature this statement includes the assumption that the starting point at high $T$ is the minimum of the MDSB-uplifted KKLT potential, which is near the origin of field space of the MDSB sector.}

We will restrict our considerations to the tree level and one-loop contributions to the
O'KKLT effective potential at nonzero temperature. So we will make use of the general finite
temperature results of \cite{BG1,BG2} for chiral multiplets coupled to supergravity at one-loop.
As is well-understood by now, the nonrenormalizability of this theory is not an issue
since it is not supposed to be viewed as a fundamental theory, but rather as an effective
low-energy description.\footnote{For
more details on one-loop computations in nonrenormalizable theories (supergravity coupled with
various matter multiplets) see \cite{OneLoop}. As was shown there for the zero-temperature
case, there are many subtleties that one has to be careful about when considering
arbitrary curved backgrounds. It is undoubtedly of great interest to achieve the same
level of understanding for $T\!\neq 0$ as well, but this goes well beyond the scope of
our paper.} However, there is another issue we should comment on, that pertains to every system that includes gravity. Namely, this is the instability under long wave-length gravitational perturbations \cite{Jeans}; it was shown in \cite{GPY} that the Jeans instability occurs also at finite temperature. While this phenomenon is of crucial importance for structure formation in the early Universe, we will not address it here and instead will limit ourselves to the
effective-potential formula of \cite{BG1,BG2}. The latter may be viewed as a good description for spatial regions of size smaller than the Jeans length\footnote{Recall, that this is the maximal length-scale for stable perturbations.} or as a necessary ingredient in a complete consideration that would take into account the above dynamical instability. Note however, that this instability is a subleading-order effect on cosmological scales, on which the Universe is well-approximated by the model of a homogeneous fluid. So, to leading order, considerations based on \cite{BG1,BG2} provide a reliable descritption of the overall thermodynamic behaviour.

The organization of this paper is the following. In Sections 2 and 3 we briefly review the O'KKLT model and the relevant properties of its zero-temperature potential. We also introduce useful notation and give a clear derivation of an order-of-magnitude relation between parameters, that is necessary for the existence of dS vacua when the non-perturbative superpotential contains a single exponent (as in the original KKLT proposal). In Section 4 we compute the temperature-dependent contribution $V_T$ to the effective potential at one loop. In Subsection 4.1 and Appendices A and B we show that $V_T$ does not have any minima at finite field vevs for the one-exponential case. In Section 5 we study the case of a race-track type superpotential with two-exponentials. We find various sets of parameters for which $V_T$ has minima at finite vevs. However, it turns out that for these parameters the minima of the total effective potential are still determined by those of the zero-temperature part, as long as the temperature is much smaller than the Planck mass (which is necessary for the reliability of the supergravity approximation). Hence, there is a regime in which the zero-temperature dS minimum of the field $\rho$ is not destabilized by thermal corrections in the supergravity approximation. In view of that, in Section 6 we reconsider the one-exponential case and find the conditions under which the minimum of $V_{eff}$ is not destabilized by the runaway temperature contribution, again for temperatures smaller than the Planck mass.

\section{O'KKLT model}

The O'KKLT model of \cite{KL} is a combination of the original KKLT model with
\be \label{KKLT}
K= - 3 \ln (\rho + \bar{\rho}) \, , \qquad W = W_0 + A e^{-a \rho}
\ee
and the O'Raifeartaigh model. The latter has three scalar fields. However, \cite{KL} considered the regime in which the two heavy ones are integrated out. One is then left with a single field $S$ with a superpotential and K\"{a}hler potential given by
\be \label{OR}
W_{O'} = -\mu^2 S \, , \qquad K_{O'} = S \bar{S} -\frac{(S \bar{S})^2}{\Lambda^2} \, .
\ee
The last term in $K_{O'}$ is due to the leading contribution in the one-loop correction in an expansion in $\frac{\lambda^2 S \bar{S}}{m^2} <\!\!< 1$, where $m$ and $\lambda$ are the remaining couplings in the full O'Raifeartaigh superpotential: $m \phi_1 \phi_2 + \lambda S \phi_1^2 -\mu^2 S$. The parameter $\Lambda$ in $K_{O'}$ denotes a particular combination of couplings, namely $\Lambda^2 = \frac{16 \pi^2 m^2}{c \lambda^4}$ with $c$ being a numerical constant of order 1. As in \cite{KL}, we will assume that $m, \mu, \Lambda <\!\!< 1$ (we work in units $M_P=1$). Also, the validity of the approximation (\ref{OR}) requires small $S$, such that $S \bar{S} <\!\!< m^2/\lambda^2 <\!\!< 1$.

In fact, as explained in \cite{KL}, for cosmological reasons it is valuable to consider a modification of (\ref{KKLT}) with two exponents:
\be \label{RaceTr}
W = W_0 + A e^{-a \rho} + B e^{-b \rho} \, .
\ee
The reason is that this race-track type superpotential, unlike (\ref{KKLT}), allows one to reconcile light gravitino mass with standard models of inflation. For convenience, from now on we will call O'KKLT model the combined system:
\be \label{OKKLT}
K = - 3 \ln (\rho + \bar{\rho}) + S \bar{S} - \frac{(S \bar{S})^2}{\Lambda^2} \, , \qquad W = W_0 + f(\rho) - \mu^2 S \, ,
\ee
where the function $f$ is either
\be \label{ftwo}
f(\rho) = A e^{-a \rho} \qquad {\rm or} \qquad f(\rho) = A e^{-a \rho} + B e^{-b \rho} \, .
\ee
Recall that, when the non-perturbative superpotential is due to gaugino condensation, the parameters in the exponents are of the form $a=2\pi/n$ and $b=2\pi/m$ with integer $n$ and $m$. In the following, however, we will only consider the effective model determined by (\ref{OKKLT})-(\ref{ftwo}), without being concerned with the microscopic physics behind it.

At zero temperature the KKLT model alone, (\ref{KKLT}), has a single AdS vacuum at finite value of $\rho$. The O'Raifeartaigh model uplifts this minimum to a positive cosmological constant vacuum. Similarly, the modified superpotential (\ref{RaceTr}) leads to two AdS vacua at finite $\rho$ and the presence of the O'Raifeartaigh field lifts one of them to dS.\footnote{Of course, the Dine-Seiberg minimum at infinity is always there too.} Our goal will be to study the phase structure of the theory (\ref{OKKLT}) at finite temperature aiming to answer the question whether the fields roll towards the dS vacuum upon cooling down. Let us first review some results about the zero temperature scalar potential, that we will need in later sections.

\section{Zero temperature potential} \label{ZTPot}
\setcounter{equation}{0}

At zero temperature the scalar potential of the system (\ref{OKKLT}) is given by the standard $N=1$ supergravity expression:
\be \label{pot}
V_0 = e^K (K^{A\bar{B}} D_A W D_{\bar{B}} \overline{W} - 3 |W|^2) \, .
\ee
One can show that its minima are obtained for vanishing imaginary parts of the scalars $\rho$ and $S$ \cite{KL}. Let us denote the real parts by $Re \rho = \sigma$ and $Re S = s$. As we only consider the moduli space region where $S$ is small, we can expand (\ref{pot}) in powers of $s$:
\be \label{ZTexp}
V_0 = V^{(0)}_0 + V^{(1)}_0 s + V^{(2)}_0 s^2 + {\cal O} (s^3) \, .
\ee
The value of $s$ at the minimum is determined by $\pd V_0 / \pd s = 0$ and so, up to ${\cal O}(s^3)$, it is \cite{KL}:
\be
s = - \frac{V^{(1)}_0}{2 V^{(2)}_0} \approx \frac{\sqrt{3}}{6} \Lambda^2 \, .
\ee

It will be useful for the future to record here the general expressions for the first three coefficients of the expansion with $K$ and $W$ as in (\ref{OKKLT}) and $\rho = \bar{\rho} = \sigma$:
\bea \label{PotExpan}
\!V^{(0)}_0 \!\!&=& \!\!\frac{\mu ^4}{8 \sigma ^3}-\frac{(3 W_0 + 3 f - \sigma f') f'}{6 \sigma ^2} \nn \\
\!V^{(1)}_0 \!\!&=& \!\!-\frac{\mu ^2 \left(W_0+f-2 \sigma f'\right)}{4 \sigma^3} \nn \\
\!V^{(2)}_0 \!\!&=& \!\!\frac{1}{8 \sigma^3} \left[\!\left( \frac{4 \mu ^4}{\Lambda ^2}+3 \mu ^4+W_0^2 \right) + 2 W_0 (f-2\sigma f') + f^2 - 4 \sigma f f' + \frac{4}{3} \sigma^2 (f')^2\right]\!,
\eea
where we have denoted $f' \equiv \pd f / \pd \sigma$. Note that the constant term in the numerator of $V_0^{(2)}$, namely $\frac{4 \mu ^4}{\Lambda ^2}+3 \mu ^4+W_0^2$, is really just $4 \mu^4 / \Lambda^2 + W_0^2$ since $\Lambda <\!\!< 1$.

Let us now take a more careful look at the cases of a superpotential with one and with two exponentials in turn. This will also enable us to introduce some useful notation.

\subsection{One exponential}

In this case, we have (\ref{OKKLT}) with
\be
f (\sigma) = A e^{-a \sigma} \, .
\ee
It was argued in \cite{KL} that a good approximation for the position of the dS vacuum is the position of the supersymmetric AdS minimum. The latter is determined by the solution of $D_{\rho} W_{KKLT} = 0$, which implies \cite{KKLT}:
\be \label{AdSvac}
W_0 = - A e^{-a \sigma} \left( 1 + \frac{2}{3} a \sigma \right).
\ee

It is easy to see that the requirement that the zeroth order, $V^{(0)}_0$, in the expansion (\ref{ZTexp}) be small and positive at the dS minimum leads to:
\be \label{W0murel}
W_0^2 \approx \mu^4 \, ,
\ee
meaning that $|W_0|$ and $\mu^2$ are of the same order of magnitude. Namely, at zeroth order the vacuum energy density is:
\be
V_0^{(0)} = V_{KKLT} + \frac{\mu^4}{(\rho + \bar{\rho})^3} \, .
\ee
Since at the AdS minimum $V_{KKLT} = -3 |W_{KKLT}|^2$ and also $\rho = \bar{\rho} = \sigma$, then the condition $V_0^{(0)}|_{min} \approx 0$ implies that
\be
3 (W_0 + A e^{-a \sigma})^2 \approx \mu^4 \, .
\ee
Substituting in the latter equation $A e^{-a \sigma}$ from (\ref{AdSvac}), we find:
\be
3 W_0^2 \left( \frac{\frac{2}{3} a \sigma}{ 1+ \frac{2}{3} a \sigma} \right)^2 \approx \mu^4 \, .
\ee
Obviously, the number $\frac{2}{3} a \sigma / (1 + \frac{2}{3} a \sigma)$ is less than $1$ for any finite $a\sigma$. In addition, it is always of order $1$ as we only consider $a \sigma > 1$ in order to have a reliable one-instanton contribution to the non-perturbative superpotential. Hence, one concludes that $W_0^2$ and $\mu^4$ have to be of the same order of magnitude in order for a dS vacuum with small cosmological constant to exist.

Now, it is clear that using $f' = -a f$ one can get rid of all derivatives in (\ref{PotExpan}). Then it is easy to notice that the parameter $a$ becomes just an overall rescaling upon introducing the new variable $x = a \sigma$. For example:
\be
V_0^{(0)} = \frac{a^3}{2 x^3} \left[ \frac{\mu ^4}{4}+ W_0 f x+f^2 \left(\frac{x^2}{3}+x\right) \right]
\ee
and hence
\be \label{DerVZ}
\frac{\pd V^{(0)}_0}{\pd \sigma} =-\frac{a^4}{x^4} \left[ Q_3(x) \, f^2 + W_0 Q_2(x) f +\frac{3 \mu^4}{8} \right] \, ,
\ee
where
\be
Q_2 (x) = x +\frac{1}{2} x^2 \, , \qquad Q_3 (x) =  x + \frac{7}{6} x^2 + \frac{1}{3} x^3  \, .
\ee
Clearly then, varying $a$ does not change the vacuum structure of the model; it only shifts the positions of the extrema along the $\sigma$-axis. We will always choose $a$ such that the minima occur for $\sigma \sim {\cal O} (100)$, so that the supergravity approximation is good.

It is also clear that the value of the parameter $A$ does not affect the vacuum structure either: one can reduce it to an overall rescaling of the scalar  potential by redefining $W_0 \rightarrow A W_0$ and $\mu^2 \rightarrow A \mu^2$. So for convenience we will set $A=1$ from now on. Thus, in this case one is left with the following essential parameters: $W_0$, $\mu$ and $\Lambda$.\footnote{Although $\Lambda$ does not occur in $V_0^{(0)}$, it is present in the total potential, see (\ref{PotExpan}).} Finally, notice that (\ref{DerVZ}) has zeros only for $W_0<0$, as all other terms in the bracket are positive definite (recall that $x\ge 0$).

\subsection{Two exponentials} \label{VZZO}

Now the function $f$ in (\ref{OKKLT}) is given by
\be
f (\sigma) = A e^{-a \sigma}+B e^{-b \sigma} \, .
\ee
Realizing that again the parameter $A$ can be reduced to an overall rescaling of the scalar potential by redefining $W_0 \rightarrow A W_0$, $\mu^2 \rightarrow A \mu^2$ and $B \rightarrow A B$, we set $A=1$ in this case too. Introducing $x=a\sigma$ as before and denoting $p\equiv b/a$, we can write $V_0^{(0)}$ of (\ref{PotExpan}) as:
\bea \label{twoexpV0}
V_0^{(0)} &=& \frac{a^3}{2 x^3} \left[e^{-2x} \!\left(\frac{x^2}{3}+x\right) \!+ B^2 e^{-2 p \,x} \!\left(\frac{p^2 x^2}{3}+ p \,x\right)\!+ B e^{-(p+1)\,x} \!\left(\frac{2 p x^2}{3}+(p+1)\,x\right) \right. \nn \\
&+& \left. W_0 \,e^{-x} \!\left( \,1 + B e^{-(p-1)x} p \,\right) x + \frac{\mu ^4}{4} \right].
\eea
Clearly, the essential parameters now are $W_0$, $B$, $p$, $\mu$ and $\Lambda$.

This potential has either one (AdS) or two extrema at finite $x$. Without uplifting, the
latter are either both AdS or one Minkowski and one AdS \cite{KLBP}. Upon adding the
uplifting sector, the first of them (i.e., the one that occurs at smaller value of $x$)
becomes dS. As before, one can again argue that the position of this extremum is very
close to the position of the original supersymmetric vacuum. And also, one again finds
that the field $S$ is stabilized at $S=\bar{S}=\frac{\sqrt{3}}{6} \Lambda^2$ \cite{KL}.

\section{One-loop effective potential at finite T} \label{VeffFT}
\setcounter{equation}{0}

In the following we will be interested in the one-loop finite
temperature effective potential for the O'KKLT model. So let us
start by recalling the general expression. It was derived first
for a renormalizable field theory in \cite{DJ}, using the
zero-temperature functional integral method of \cite{Jackiw}, and
later generalized to supergravity in \cite{BG1}. Namely, the
one-loop effective potential is given by: \be V_{eff} (\hat{\chi})
= V_{tree} (\hat{\chi}) + V_0^{(1-loop)} (\hat{\chi}) +
V_T^{(1-loop)} (\hat{\chi}) \, , \ee where $V_{tree}$ is the
classical potential, $V_0^{(1-loop)}$ is the zero temperature
one-loop contribution, encoded in the Coleman-Weinberg formula,
and finally the temperature-dependent correction is: \be
\label{Texpansion} V_T^{(1-loop)}(\hat{\chi}) = -\frac{\pi^2
T^4}{90} \left( g_B+ \frac{7}{8} g_F \right)+ \frac{T^2}{24}
\left[{\rm Tr} M_s^2(\hat{\chi}) + {\rm Tr} M_f^2(\hat{\chi})
\right] + {\cal O}(T) \, . \ee Here $\{ \chi^A \}$ denotes
collectively all fields in the theory and the quantities ${\rm Tr}
M_s^2$ and ${\rm Tr} M_f^2$ are traces over the mass matrices of
all scalar and fermion fields respectively in the classical
background $\{\hat{\chi}^A\}$.\footnote{In (\ref{Texpansion}),
${\rm Tr} M_f^2$ is computed summing over Weyl fermions.} They are
given by \cite{BG2}: \be \label{TrMs} {\rm Tr} M^2_s = 2 \,\la
K^{C\bar{D}} \,\frac{\partial^2 V_0}{\partial \chi^C
\partial \bar{\chi}^{\bar{D}}} \,\ra \, ,
\ee
where $V_0$ is as in (\ref{pot}), and
\be \label{TrMf}
{\rm Tr} M_f^2 = \la e^G \left[ K^{A \bar{B}} K^{C \bar{D}} (\nabla_A G_C
+ G_A G_C) (\nabla_{\bar{B}} G_{\bar{D}} + G_{\bar{B}}
G_{\bar{D}}) - 2 \right] \ra \, ,
\ee
where $G = K + \ln |W|^2$. The constants $g_B$ and $g_F$ in (\ref{Texpansion}) are the
total numbers of bosonic and fermionic degrees of freedom. As in the following we will
only be interested in derivatives of $V_{eff}$ w.r.t. some $\hat{\chi}^A$, we will drop
from now on the $T^4$ piece of the effective potential. Also, for convenience, we will
denote the remaining expression in (\ref{Texpansion}) just by $V_T$. We should also note
that the high temperature expansion (\ref{Texpansion}) is only valid in the regime, in
which all masses are much smaller than the energy scale set by the temperature.

We will compute the effective potential in the classical background:
\be \label{BGR}
\la \rho \ra = \langle \bar{\rho} \rangle = \sigma \, , \qquad \langle S
\rangle = \langle \bar{S} \rangle = s \, .
\ee
As one can notice, we are using the same notation for the above vevs as for the real
and imaginary parts of the fields $\rho$ and $S$. This is convenient since the classical
potential in the expression for $V_{eff}$ is always understood to be evaluated in
the background $\hat{\chi}$, which in our case is (\ref{BGR}). This abuse of notation
should not cause any confusion; it will always be clear from the context what we mean.

In the rest of this section we will concentrate on the temperature-dependent part, $V_T$, of the effective potential. The reason is that at high temperature it is expected that $V_T$ dominates the behaviour of $V_{eff}$. We will be more detailed on this in Section \ref{PhSFT}.
As before, we will confine our considerations to the region of field space, in which $s$ is
small. So, similarly to (\ref{ZTexp}), we can expand the temperature-dependent part of
the effective potential as:
\be \label{VTexp}
V_T = V_T^{(0)} + V_T^{(1)} s + V_T^{(2)} s^2 + {\cal O} (s^3) \, .
\ee
Applying formulae (\ref{Texpansion})-(\ref{TrMf}) for $K$ and $W$ given by
(\ref{OKKLT}), one finds:
\bea \label{V2T}
V^{(0)}_T &=& \frac{T^2}{24} \,\frac{1}{\sigma^3} \left[ \frac{\mu^4}{\Lambda^2} + \frac{1}{2} W_0^2 + W_0 \left( f-\frac{13}{3} \sigma f' + 2 \sigma^2 f'' \right) \right. \nn\\
&+& \left. \frac{f^2}{2} -\frac{13}{3} \sigma f f' + \sigma^2 \left( \frac{25}{9} (f')^2 + 2 f f'' \right) - \frac{8}{3} \sigma^3 f' f'' + \frac{2}{3} \sigma^4 (f'')^2 \right]\!, \nn \\
V^{(1)}_T &=& - \frac{T^2}{12} \,\frac{\mu^2}{\sigma^3} \left[ \frac{1}{\Lambda^2} (W_0 + f) - \frac{11}{3} \sigma f' + \sigma^2 f'' \right]\!, \nn \\
V^{(2)}_T &=& \frac{T^2}{24} \,\frac{1}{\sigma^3} \left[ 22 \,\frac{\mu^4}{\Lambda^4} + \frac{W_0^2}{\Lambda^2}
+ W_0 \left( \frac{2}{\Lambda^2} f -\frac{22}{3} \sigma f' + 2 \sigma^2 f'' \right) \right. \nn \\
&+& \left. \!\frac{1}{\Lambda^2} f^2 -\frac{22}{3} \sigma f f' + \sigma^2 \left( \frac{34}{9} (f')^2 + 2 f f'' \right) - \frac{8}{3} \sigma^3 f' f'' + \frac{2}{3} \sigma^4 (f'')^2 \right]\!.
\eea
Here, in expressions like $3+1/\Lambda^2$ we have kept only the last term since $\Lambda^2 <\!\!<1$.
Also, it is understood that $f' \equiv \la \,\pd \!f / \,\pd \rho \ra$ and
$f'' \equiv \la \,\pd^2 \!f / \,\pd \rho^2 \ra$.\footnote{Obviously
$\la \,\pd \!f / \,\pd \rho \ra = \la \,\pd \!\bar{f} / \,\pd \bar{\rho} \ra$.} Note
that, similarly to Section \ref{ZTPot}, the value of $s$ at the minimum is
\be \label{sVT}
s_{min} = - \frac{V_T^{(1)}}{2 V_T^{(2)}}\bigg|_{\sigma=\sigma_{min}} \,\,\, ,
\ee
up to ${\cal O}(s^3)$ in $V_T$.

Let us now take a more careful look at the temperature-dependent effective potential $V_T$
for each of the two cases in (\ref{ftwo}). We will concentrate on the leading contribution
$V_T^{(0)}$. At the end we will check that (\ref{sVT}) gives $|s_{min}|<\!\!< 1$,
as in the zero-temperature case, and so the zeroth order of the $s$-expansion in
(\ref{VTexp}) is indeed a good approximation for $V_T$.

\subsection{One exponential} \label{OneExp}

We consider first
\be
f (\rho) = A e^{-a \rho} \quad , \,\, {\rm i.e.} \qquad W = W_0 + A e^{-a \rho} -\mu^2 S \, .
\ee
Similarly to the $T=0$ case, one can use $f' = -a f$ and $f'' = a^2 f$ to get rid of
all derivatives. Then, after introducing the variable $x = a \sigma$, one has:
\be
V^{(0)}_T = \frac{T^2}{24} \,\frac{a^3}{x^3} \left[ P_4(x)  f^2  + W_0 P_2(x) f + C_0 \right],
\ee
where
\bea \label{P2P4}
P_4(x) &=& \frac{2 x^4}{3}+\frac{8 x^3}{3}+\frac{43 x^2}{9}+\frac{13 x}{3}+\frac{1}{2} \, , \nn \\
P_2(x) &=& 2 x^2+\frac{13 x}{3}+1 \, ,
\eea
and
\be
C_0 \equiv \frac{\mu^4}{\Lambda^2} + \frac{W_0^2}{2} \, .
\ee
So the extrema are determined by:
\be \label{ExtremaVT}
\frac{\pd V^{(0)}_T}{\pd \sigma} = - \frac{T^2}{24} \, \frac{a^4}{x^4} \!\left(C + P_5(x) e^{-2 x} + W_0 P_3(x) e^{-x}\right) \!=0 \, ,
\ee
where $C \equiv 3 C_0$ and
\bea
P_5(x) &=& \frac{3}{2} + \frac{29}{3} x + \frac{121}{9} x^2 + \frac{86}{9} x^3 + \frac{14}{3} x^4 + \frac{4}{3} x^5 \, , \nn \\
P_3(x) &=& 3 + \frac{29}{3} x +\frac{19}{3} x^2 + 2 x^3 \, .
\eea

As in the zero temperature case, the constant $a$ is just an overall rescaling which does
not affect the presence or absence of minima. Also, the constant $A$ can again be reduced
to an overall rescaling by redefining $W_0 \rightarrow A W_0$ and
$\mu^2 \rightarrow A \mu^2$. So again we can set $A=1$ without loss of generality.
Of course, this is not a surprise; as we are describing the same system at $T=0$ and
$T\neq 0$, we should have the same parameters in both cases.

\begin{figure}[t]
\begin{center}
\hspace{-0.3cm} \scalebox{0.8}{\includegraphics{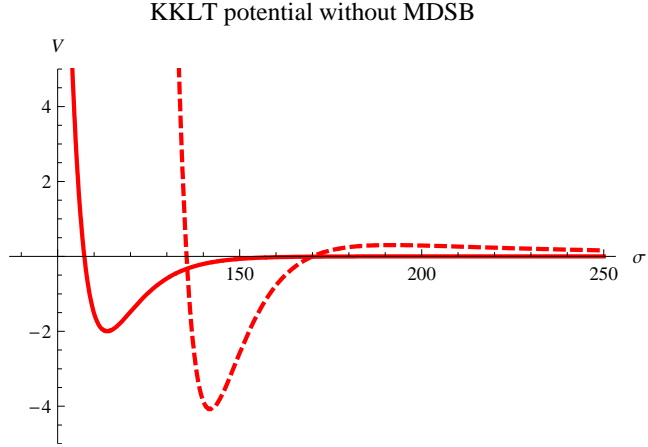}}
\end{center}
\vspace{-0.5cm}
\caption{ {\small O'KKLT potential (multiplied by $10^{15}$) for the one exponential case without MDSB sector, i.e. simply KKLT potential, for the parameter values: $W_0 = - 10^{-4}$ and $a=0.1$. One can see that, unlike the case with MDSB sector, both the zero temperature potential (red continuous line) and the finite temperature part (red dashed line) have a minimum for finite vev of $\sigma$. (For convenience, the graph of the temperature-dependent part is actually a plot of $V_T / (\frac{T^2}{24})$ vs $\sigma$.})}
\label{OneExpWmsdb1}
\end{figure}

Looking at (\ref{ExtremaVT}), one immediately sees that an obvious minimum of $V_T^{(0)}$
is obtained for $x\rightarrow \infty$. However, we are interested in solutions at finite
field vevs. In other words, we would like to solve
\be
C + P_5(x) e^{-2 x} + W_0 P_3(x) e^{-x} = 0 \, ,
\ee
or equivalently:
\be \label{E1t}
e^{2 x} = - C^{-1} (P_5 + W_0 P_3 e^x) \equiv H(x) \, .
\ee
Unfortunately, this equation cannot be solved analytically. Clearly though,
its solutions (and, in fact, the presence or absence of such) depend(s) on the values of
the parameters $C$ and $W_0$. In Appendix B we use the
method described in Appendix A in order to show that (\ref{E1t}) {\it does not} have any
solutions for parameters such that the zero-temperature potential has a dS minimum at
finite $x$ (equivalently, at finite $\sigma$). This is not trivial as removing the
restriction for the existence of a dS minimum at $T=0$, for instance by taking $\mu=0$
and so completely turning off the uplifting, one finds that $V_T^{(0)}$ {\it can} have a
minimum at finite $\sigma$. See Figure \ref{OneExpWmsdb1} for an explicit example.

\subsection{Two exponentials} \label{VTZO}

Now we turn to the second case in (\ref{ftwo}). Namely, we consider
\be
f (\rho) = A e^{-a \rho} + B e^{-b \rho} \, .
\ee
In terms of $x=a\sigma$ and $p=b/a$ one has from (\ref{V2T}):
\bea \label{twoexpVT}
V_T^{(0)} &=& \frac{T^2}{24} \,\frac{a^3}{x^3} \left[e^{-2 x} P_4(x) + B^2 e^{-2 p \,x} P_4(p \,x) + B e^{-(p+1)x} \,Q_4(x) \right. \nn \\
&+& \left. W_0 \left( e^{-x} P_2(x)+e^{-p\,x} P_2(p\,x) \,\!\right) + \,C_0 \right],
\eea
where the polynomials $P_2$, $P_4$ were defined in (\ref{P2P4}) and
\be
Q_4(x) = P_2(x)+P_2(p \,x) - 1 + \frac{2 p \,x^2}{3} \,\left( 2 p \,x^2 +4(1 + p)x +25 \right) \, .
\ee

\begin{figure}[t]
\begin{center}
\scalebox{0.7}{\includegraphics{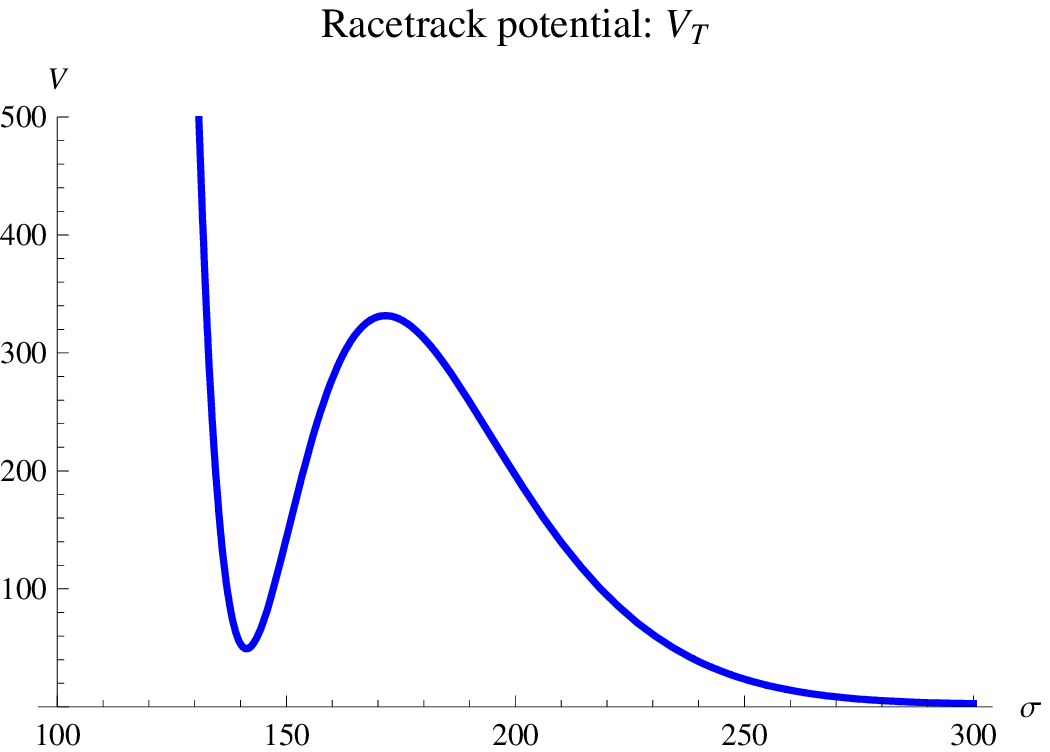}} \scalebox{0.7}{\includegraphics{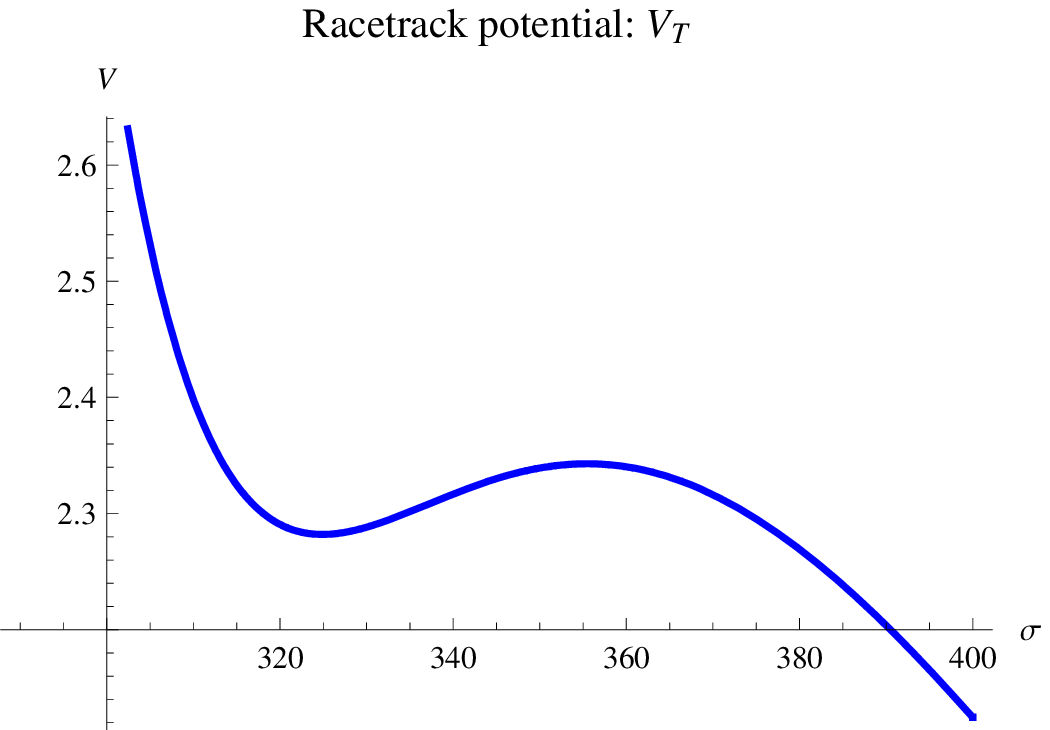}}
\end{center}
\vspace{-0.5cm}
\caption{ {\small Temperature-dependent part of the effective potential (multiplied by $10^{15}$) in the case of a race-track type superpotential for parameter values: $a=\pi/100$, $B =-1.028$, $p=\frac{100}{99}$, $W_0=-2.4\times 10^{-4}$, $\mu = 0.66 \times 10^{-3}$, $\Lambda=10^{-3}$. One can see that there are
two local minima at finite $\sigma$; note also that the second one (right) is  significantly shallower than the first (left). As in Figure \ref{OneExpWmsdb1}, for convenience we have actually plotted $V_T / (\frac{T^2}{24})$ vs $\sigma$.}}
\label{OKLVT}
\end{figure}

Unlike in the previous subsection, $V_T^{(0)}$ of (\ref{twoexpVT}) {\it can} have finite
$\sigma$ minima for parameters, for which $V_0$ has a dS vacuum. This is examplified in
Figure \ref{OKLVT} for a particular choice of parameters. Thus, in the next section we will concentrate on investigating the phase structure of the system for the case of two exponentials in the non-perturbative superpotential.

\section{Phase structure at finite T} \label{PhSFT}
\setcounter{equation}{0}

In this section we study the finite temperature phase structure of the O'KKLT model with two exponents in the superpotential. Let us first summarize the necessary ingredients of the set-up and the questions we would like to address.

As we reviewed in Subsection \ref{VZZO}, the zero temperature potential generically has two minima with finite field vevs. The same is true also for $V_T^{(0)}$ (see Figure \ref{OKLVT}). One naturally expects that at high temperature the system is in a local
minimum of the temperature-dependent part of the effective potential. We assume that this
starting point is the lower-$x$ minimum of $V_T^{(0)}$.\footnote{This is preferable than
the other finite-$x$ minimum, as the latter one is much shallower. Of course, if
one were to start from the global minimum, which is at infinity, then clearly there would
be nothing to discuss $-$ the system would remain there at any temperature and so one would
have the undesirable situation of decompactified internal space at $T\!=0$.} Then, as the
temperature decreases, a point will be reached at which a second order phase transition
will occur and the system will start rolling towards one of the zero temperature minima.
The critical temperature $T_c$ for this to happen, as well as the relevant field space position $x_c$, can be found by solving the following set of equations:
\be \label{Tcxc}
V'_{eff} (T_c, x_c) = 0 \qquad {\rm and} \qquad V''_{eff} (T_c, x_c) = 0 \, ,
\ee
where $'$ denotes $d/dx$. We would like to know whether as a result of this phase transition the system will start rolling towards the zero-temperature dS vacuum or in the opposite direction, i.e. towards the $T \!\!= \!0$ supersymmetric vacuum.\footnote{As we will see below, the finite $T$ minimum, that is our starting point, is always between the dS and the susy $T\!=0$ vacua.} Clearly, this would be determined by the sign of $V_{eff}'''(T_c,x_c)$.

Note that for systems with $x\rightarrow -x$ symmetry, as is the
case for the ISS model for example (see \cite{FKKMT}), the origin of field space is a
local minimum of $V_{eff}$. If one takes this as the starting point at high $T$, then the
first equation in (\ref{Tcxc}) is identically satisfied. So, from the second equation, one
is left with the familiar
\be
V_T'' = - m^2
\ee
as the condition that determines $T_c$ of a second order phase transition. Recall
that $m^2$ here is the classical mass, clearly originating from $V_0$, of the scalar field with nonzero background vev. In our case however, none of the equations in (\ref{Tcxc}) is trivial and so we have to solve simultaneously both of them.

To facilitate our considerations, let us change variables to the real components of the fields: $\rho= Re \rho + i Im \rho$, $\bar{\rho}= Re \rho - i Im \rho$. As
in the background (\ref{BGR}) one has $\la Re \rho \ra = \sigma$ and $\la Im \rho \ra = 0$, clearly the field $Re \rho$ is the one that drives the phase transition we are after. So (\ref{Tcxc}) acquires the form:
\be \label{Tcxc2}
\frac{\pd V_T}{\pd x} = - \frac{\pd V_0}{\pd x} \, , \qquad \frac{\pd^2 V_T}{\pd x^2} = - m^2_{Re \rho} \, ,
\ee
where in the second equation we have used that $\pd^2_x V_0 (x) = m^2_{Re \rho} (x)$. One can compute the first and second derivatives of $V_T$ and $V_0$ using (\ref{twoexpVT}) and (\ref{twoexpV0}), respectively. Before turning to that however, let us first show that the variables $Re \rho$ and $Im \rho$ diagonalize the classical mass matrix.

\subsection{Mass matrix}

From (\ref{pot}) one can easily find the tree-level mass-squared matrix in the background (\ref{BGR}).\footnote{As before, we only consider the zeroth order in the $s$-expansion.} It is of the form:
\be
M = \left( \begin{array}{cc} m^2_{\rho \rho}  & m^2_{\rho \bar{\rho}} \\ m^2_{\bar{\rho} \rho} & m^2_{\bar{\rho} \bar{\rho}} \end{array} \right),
\ee
where \,$m^{2}_{\rho \bar{\rho}} \equiv \la \pd_{\rho} \pd_{\bar{\rho}} V_0 \ra = \pd_{\rho} \pd_{\bar{\rho}} V_0|_{\rho =\bar{\rho} = \sigma, \,S=\bar{S}=0}$ \,etc. with all matrix elements being nonzero and $m^2_{\rho \rho} = m^2_{\bar{\rho} \bar{\rho}}$\,, \,$m^2_{\rho \bar{\rho}} = m^2_{\bar{\rho} \rho}$. It is clear then, that the real components of $\rho$ have diagonal mass matrix with masses:\footnote{Obviously, the matrix $M$ is diagonalized by the change of variables $\rho_+ = \frac{\rho + \bar{\rho}}{\sqrt{2}}$ and $\rho_- = \frac{\rho - \bar{\rho}}{\sqrt{2}}$ with the corresponding eigenvalues being $m^2_{\rho_{\pm}} \!\!= m^2_{\rho \rho} \pm m^2_{\rho \bar{\rho}}$. On the other hand, $\rho_+ = \sqrt{2} \,Re \rho$ and so $m^2_{Re \rho} = 2 m^2_{\rho_+}$; similarly $m^2_{Im \rho} = -2 m^2_{\rho_-}$.}
\be
m^2_{Re \rho} = 2 \left(m^2_{\rho \bar{\rho}}+m^2_{\rho \rho} \right)\, , \qquad m^2_{Im \rho} = 2 \left(m^2_{\rho \bar{\rho}}-m^2_{\rho \rho} \right)\, .
\ee

In terms of a generic function $f(\rho)$ in (\ref{OKKLT}), these expressions are:
\bea \label{Mp}
m^2_{Re \rho} \!\!\!&=& \!\!\frac{1}{6 \sigma^5}\left[ 9 \mu ^4+ 3 W_0 \left( -6 \sigma f' + 4 \sigma^2 f'' -\sigma^3 f^{(3)}\right) \right. \\
\!\!\!&-& \!\!\!\left. 18 \sigma f f' + 2\sigma^2 \left( 7 (f')^2 + 6 f f'' \right) - \sigma^3 \left(13 f' f'' + 3 f f^{(3)}\right)  + 2 \sigma^4 \left( (f'')^2+f' f^{(3)}\right) \right] \nn
\eea
and
\be \label{Mm}
m^2_{Im \rho} = \,\frac{1}{6 \sigma^5} \left[ 3 W_0 \,\sigma^3 f^{(3)} - 3 \sigma^3 \left(f' f'' - f f^{(3)}\right) +2 \sigma^4 \left( (f'')^2 - f' f^{(3)} \right) \right].
\ee
Specializing (\ref{Mp}) and (\ref{Mm}) to the case of two exponents and introducing $x=a \sigma$ and $p=b/a$ as before, we can also write:
\bea \label{Mpl}
m^2_{Re \rho} &=& \frac{a^5}{3 x^5} \left[e^{-2x} R_4(x) + B^2 e^{-2p\,x} R_4(p\,x) + B e^{-(p+1)x} S_4(x) \right. \nn \\
&+& \left. W_0 \left( e^{-x} R_3(x)+ B e^{-p\,x} R_3(p\,x) \right) + \frac{9 \mu ^4}{2} \right],
\eea
where
\bea
R_4 (x) &=& 2 x^4+8 x^3+13 x^2+9 x \, , \qquad R_3 (x) = \frac{3 x^3}{2}+6 x^2+9 x \, , \nn \\
S_4 (x) &=& R_3(x) + R_3(p\,x) + px^2 \!\left[ 14 + (p+1)x \!\left( \!\!(p+1)x + \frac{13}{2}\!\right) \!\right],
\eea
and
\be \label{Mmi}
m^2_{Im \rho}= - \frac{a^5}{2 x^5} \left[ B e^{-(p+1) x} (p-1)^2 x^2 \!\left( \!(p+1) x + \frac{2}{3} p\,x^2 \!\right)+ W_0 (e^{-x} + B e^{-p\,x} p^3) x^3 \right]\!.
\ee
Recall that $W_0 < 0$ and so, despite the overall minus in the above formula, $m^2_{Im \rho}$ is not negative definite.

\subsection{Critical temperature}

Let us now turn to solving (\ref{Tcxc2}) in order to find $T_c$. As we already mentioned, one can compute the relevant derivatives of $V_0$ and $V_T$ from (\ref{twoexpV0}) and (\ref{twoexpVT}). However, the resulting equations are of the same type as (\ref{E1t}), only significantly more complicated. Thus, one cannot hope to analyze them analytically. So we will study them numerically for various values of the parameters.

We should note first, that clearly there are many parameter values for which the system does not exhibit the behaviour we described in the beginning of this section. Namely, it could happen that even though $V_0$ has a dS vacuum, the only minimum that $V_T$ has is the Minkowski one at $\la \rho \ra = \infty$; or it could be that, instead of two minima at finite $\la \rho \ra$, $V_T$ has only one. This is similar to the situation at zero-temperature: There are many values of the parameters for which $V_0$ does not have a dS minimum. The important point, however, is that there are also many values for which a dS vacuum does exist at $T=0$ \cite{KL}; they are exactly the parameter values of interest in the search for moduli stabilized dS vacua. Similarly, here we concentrate on the regime for which $V_T$ does have at least one finite-vevs minimum {\it when} $V_0$ has a dS vacuum. The corresponding  choices of parameters are exactly those for which the internal space of a dS compactification has the chance of not being destabilized by thermal effects. And that possibility is precisely what we want to explore.

Several sets of parameters, for which the system is in the desired regime, are given in Table \ref{table}.
\begin{table}[h]
\begin{center}
\vspace{0.4cm}
\begin{tabular}{|c|c|c|c|c|c|c|c|c|}
\hline
 $B$ & $W_0$ & $\mu$  & $\Lambda$ & $x_{dS}^{(0)}$ & $x_{AdS}^{(0)}$ & $x_{min}^{(T)}$ & $x_c$ & $T_c$ \\
\hline
-1.040 & $-7.6 \times 10^{-5}$ & $8 \times 10^{-4}$ & $10^{-2}$ & 4.88 & 7.84 & 5.62 & 5.54 & 0.27 \\
\hline
-1.036 & $-1.1 \times 10^{-4}$ & $2 \times 10^{-3}$ & $10^{-2}$ & 4.50 & 7.40 & 5.25 & 5.10 & 0.27 \\
\hline
-1.032 & $-1.64 \times 10^{-4}$ & $ 10^{-3}$ & $10^{-2}$ & 4.11 & 6.92 & 4.83 & 4.73 & 0.30 \\
\hline
-1.028 & $-2.4 \times 10^{-4}$ & $0.66 \times 10^{-3}$ & $10^{-3}$ & 3.73 & 6.44 & 4.44 & 4.31 &  0.30 \\
\hline
-1.024 & $-3.533 \times 10^{-4}$ & $0.66 \times 10^{-3}$ & $10^{-3}$ & 3.34 & 6.00 & 4.04 & 3.91 & 0.33 \\
\hline
-1.020 & $-5.21 \times 10^{-4}$ & $0.95 \times 10^{-3}$ & $10^{-3}$ & 2.96 & 5.52 & 3.64 & 3.47 & 0.34 \\
\hline
-1.016 & $-7.67 \times 10^{-4}$ & $1.4 \times 10^{-3}$ & $10^{-2}$ & 2.55 & 5.02 & 3.20 & 3.08 & 0.38 \\
\hline
\end{tabular}
\end{center}
\vspace{-0.2cm}
\caption{ {\small Each row of this table represents a set of parameters for which both $V_0$ and $V_T$ have minima at finite field vevs {\it and} the lower-$x$ minimum of $V_0$ is dS. In each set $p=100/99$ as in the examples of \cite{KL}. The positions of the minima are denoted by $x_{dS}^{(0)}$ and $x_{AdS}^{(0)}$ for $V_0$ and by $x_{min}^{(T)}$ for the lower-$x$ minimum of $V_T$. Recall also that $x=a\sigma$ and so, taking $a=\frac{\pi}{100}$ for instance,
the various minima occur for $\sigma \sim {\cal O} (100)$.} \label{table}}
\end{table}
This table suggests that the O'KKLT model exhibits the behaviour, that we want to study,
only at discrete points in parameter space. However, this is not completely true: for
some of the sets one can vary somewhat one (or more) parameter(s) without exiting the
regime of interest.\footnote{For example, in the first row $\mu$ can be anything between
$8\times 10^{-4}$ and $1\times 10^{-3}$; in the third row $\Lambda$ can also be $10^{-3}$;
another variation of the third row is for instance $B= -1.031$, $W_0 = -1.8 \times 10^{-4}$,
$\mu= 10^{-3}$, $\Lambda= 10^{-2}$ or $10^{-3}$; in the fifth row $W_0$ can be anything
between $-3.532 \times 10^{-4}$ and $-3.535 \times 10^{-4}$; in the seventh row $\Lambda$
can be anything between $10^{-3}$ and $10^{-2}$ etc..} Nevertheless, it is an
interesting observation that a more significant change of one parameter (with the exception of $\Lambda$) seems to require
such a change in at least one other parameter. This pattern is there, regardless of the
(runaway or not) behaviour of $V_T$, as long as one looks for parameter values giving dS
vacua of the zero temperature potential studied in \cite{KL}. Hence, the O'KKLT model
may be an example of how arguments of the kind of \cite{SW} might fail. Namely, in those
arguments one usually varies a single constant of nature (the cosmological constant, for
instance) while keeping all other coupling constants fixed. And one concludes that such
variations lead to drastic changes in the resulting physics. However, it might be that in
order to get to a new background, that is quite similar to the original vacuum, one has to
change in a discrete way (as opposed to varying continuously) more than one constant of nature at the same time. It is conceivable then, that the
above-mentioned anthropic/environmental arguments for the value of the cosmological
constant could break down under such more general variations.\footnote{See however \cite{Hamed} for arguments in favor of Weinberg's argument in the case when only the cosmological constant and the Higgs mass are varied.} This is certainly worth
investigating in more depth and within more realistic models; we hope to come back to
it in the future.

\begin{figure}[t]
\begin{center}
\scalebox{0.72}{\includegraphics{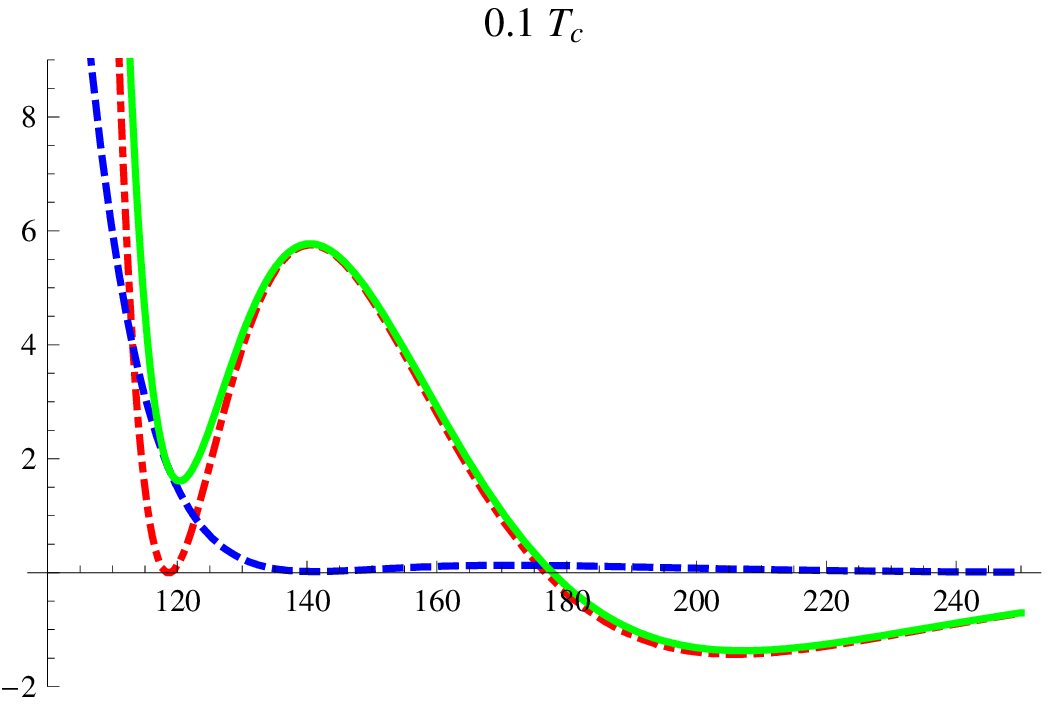}} \hspace*{0.4cm}\scalebox{0.72}{\includegraphics{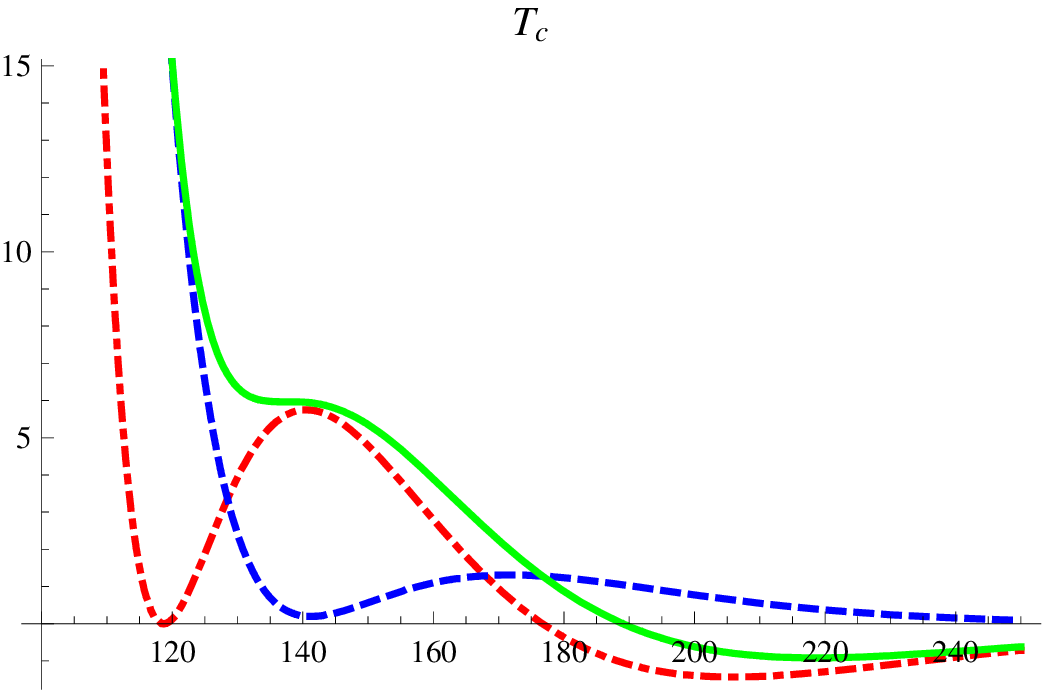}}
\end{center}
\vspace{-0.5cm}
\caption{ {\small Effective potential $V_{eff}$ (green continuous line), multiplied by $10^{15}$, as a function of $\sigma$ for the race-track type model compared to $V_0 (\sigma)$ (red dot-dashed line) and $V_T (\sigma)$ (blue dashed line) for $T=0.1 \times T_c$ (left) and for
$T= T_c$ (right). The values of the parameters are the following: $a=\pi/100$, $B = -1.028$, $p=\frac{100}{99}$, $W_0=-2.4\times 10^{-4}$, $\mu = 0.6
\times 10^{-3}$, $\Lambda=10^{-3}$; the resulting critical temperature is $T_c = 0.3$.}}
\label{OKLTc}
\end{figure}
As one can see from Table \ref{table}, the $V_T$ minimum of
interest is always between the zero temperature dS and AdS vacua.
So it might seem that a meaningful question to ask is whether the
system rolls towards the metastable or the supersymmetric $T=0$
vacua as it cools down. However, the critical temperature for the
relevant second order phase transition turns out to be always of
order $0.1$ (see Table \ref{table}; an example is illustrated in
Figure \ref{OKLTc}\footnote{The graph of $V_{eff}$ (the green
continuous  line) on this figure still does not include the term
$\sim T^4$ in (\ref{Texpansion}); as the latter is
$x$-independent, it only leads to an irrelevant for us overall
shift of the $V_{eff}$ plot down the vertical axis, which just
makes it rather inconvenient to illustrate the main features of
the remaining graphs on the same figure.}). Since we work in units
in which $M_P =1$, this means that $T_c \sim {\cal O}(0.1 \,M_P)$.
For such a high temperature the supergravity approximation is not
reliable anymore and so we cannot make any statement about the
occurrence or not of a phase transition.\footnote{We should note
that the Planck scale is the only cut-off of concern as long as
one is studying the field theory defined by (\ref{OKKLT}) on its
own, which is the viewpoint we take here. However, if one wants to
view it necessarily as the low-energy effective description of the
O'Raifeartaigh model that is obtained by integrating out the two
heavy fields, which was the original motivation to think about it,
then there is another lower cut-off. Namely, this is the scale at
which the single field approximation to the full O'Raifeartaigh
model stops being valid.} Nevertheless, turning this around, we
can conclude that for the whole range of validity of the
supergravity approximation (i.e., for $T <\!\!<
M_P$)\footnote{Clearly then, this is even more so for the whole
range of validity of the single field low-energy approximation to
the full O'Raifeartaigh model.} the extrema of the effective
potential of the system are determined by the zero-temperature
part and not by $V_T$.\footnote{Note that, because of the constant
term $\sim T^4$ that we are omitting, this does not mean that the
magnitude of $V_{eff}$ itself is determined by $V_0$.} Hence, at
the level of supergravity the $T=0$ de Sitter minimum is not
destabilized by thermal effects. This is quite unexpected, not
only because it turned out that $V_T$ can have minima at finite
field vevs (as opposed to having runaway behaviour), but also
because the potential barrier in $V_{eff}$, that separates those
minima from the $T=0$ ones, is many orders of magnitude smaller
than the Planck scale (more precisely, it is $\sim 10^{-15}$; see
Figure \ref{OKLTc} which is a typical representative for all rows
of Table \ref{table}) and so intuitively one might have expected
thermal fluctuations to get the system over it at a temperature
$<\!\!< M_P$.

The above behaviour could have implications for the early Universe, if one views the O'KKLT model as a simple toy model for the latter. Namely, at the end of the inflationary stage the universe is very cold and so it would be in a local minimum of $V_0$.\footnote{Note that the
volume modulus $\sigma$ {\it should not} be confused with the inflaton field.} Let us assume that
this is the lower-$x$ dS vacuum. Now, the exit from inflation comes with the decay of the
inflaton into various other particles and the subsequent reheating of the universe to
some temperature $T_R$. If $T_R<\!\!< M_P$, as one expects in many phenomenological
models, then after reheating the system will still be in the metastable minimum; in other
words, the dS vacuum will not be destabilized by the thermal corrections.

Having in mind this new perspective, one may wonder whether the dS vacuum remains a local
minimum of $V_{eff}$ even when the temperature-dependent part of the potential does not have
other minima except the runaway one at infinity (of course, as long as $T <\!\!< M_P$).
Indeed, our interest in finite-$\sigma$ minima of $V_T$ was stemming from the expectation
that their presence would be the obstacle for decompactification of the internal space
(whose volume $\sigma$ is proportional to). However, as we saw above, in the
range of validity of our considerations this obstacle turned out to be different. So it is
a legitimate question to ask whether the minima of $V_0$ determine the minima of
$V_{eff}$ even for parameter values for which $V_T$ has runaway behaviour. One can easily
check that this is indeed the case for sets of parameters that are close to those in Table
\ref{table}, but such that $V_0$ still has a dS vacuum while $V_T$ does not have any
finite-$x$ minima. We leave for the future a more detailed investigation of this issue for
a broader range of parameter values.

Before concluding this section, let us note that one can easily verify from (\ref{Mpl}) and (\ref{Mmi}) that
$m^2_{Re \rho}$ and $m^2_{Im \rho}$ are quite small for all sets of parameters in Table
\ref{table}. That is, there is an appreciable interval for the temperature $T$, given by
$m_{Re \rho}, m_{Im \rho} <\!\!< T <\!\!< M_P$, in which the high-temperature
expansion (\ref{Texpansion}) is well-justified. Clearly, if the minima of the effective potential are
determined by $V_0$ (instead of by $V_T$) in this interval, they will remain determined by
$V_0$ at lower temperatures as well. Finally, one can also check from (\ref{sVT}) that,
similarly to the zero-temperature case, the potential $V_T (x,s)$ stabilizes the variable
$s$ at a value $|s| <\!\!< 1$ and so the leading term in the small-$s$ expansion is
indeed a good approximation for the full expression.

\section{One exponential revisited}
\setcounter{equation}{0}

As we saw in the previous section, in the case of two exponents the stabilization of the
zero temperature dS vacuum is not due to the presence of a local minimum of $V_T$ at
finite field vevs. Rather, it comes from the fact that the minima of $V_{eff}$
are determined by the $T=0$ contribution, and not by the temperature-dependent one, even at high
$T$, as long as $T<\!\!<M_P$. Given that, it is worth to re-examine the one-exponential
case in order to see whether there is a range of parameter values for which the same thing
happens in this case too.

Here again we do not include in $V_T$ the moduli-independent $T^4$ term. Since the
$T$-dependent contribution to the effective potential has runaway behaviour (see
Section \ref{OneExp}), the dS minimum of $V_0$ is not completely washed out in the total
potential only when the magnitude of $V_T$ is smaller than the magnitude of $V_0$ at the
position of this minimum $x_{dS}$. To estimate the order of magnitude of the ratio
$|V_T^{(0)}|/|V_0^{(0)}|$ at $x_{dS}$, recall that for the case of one exponential one
has the relation $\mu^2 \approx |W_0|$, see (\ref{W0murel}). Also, at $x_{dS}$ equation
(\ref{AdSvac}) holds to a good degree of accuracy. Using these, we find the order of
magnitude estimate:
\be \label{VTV0R}
\frac{|V^{(0)}_T|}{|V^{(0)}_0|} \Big|_{x=x_{dS}} \approx \left( \frac{4 x_{dS}^2+12 x_{dS}+9}{8 x_{dS}^2-12 \,x_{dS}-9} \,\,\frac{1}{\Lambda^2}
+ \frac{6 x_{dS}^2+12 x_{dS} + 1}{8 x_{dS}^2-12 x_{dS}-9} \,\,x_{dS}^2 \right) \!\frac{T^2}{3} \,\, ,
\ee
where we have expressed $\mu$ and $W_0$ in terms of $x_{dS}$.

\begin{figure}[t]
\begin{center}
\scalebox{0.72}{\includegraphics{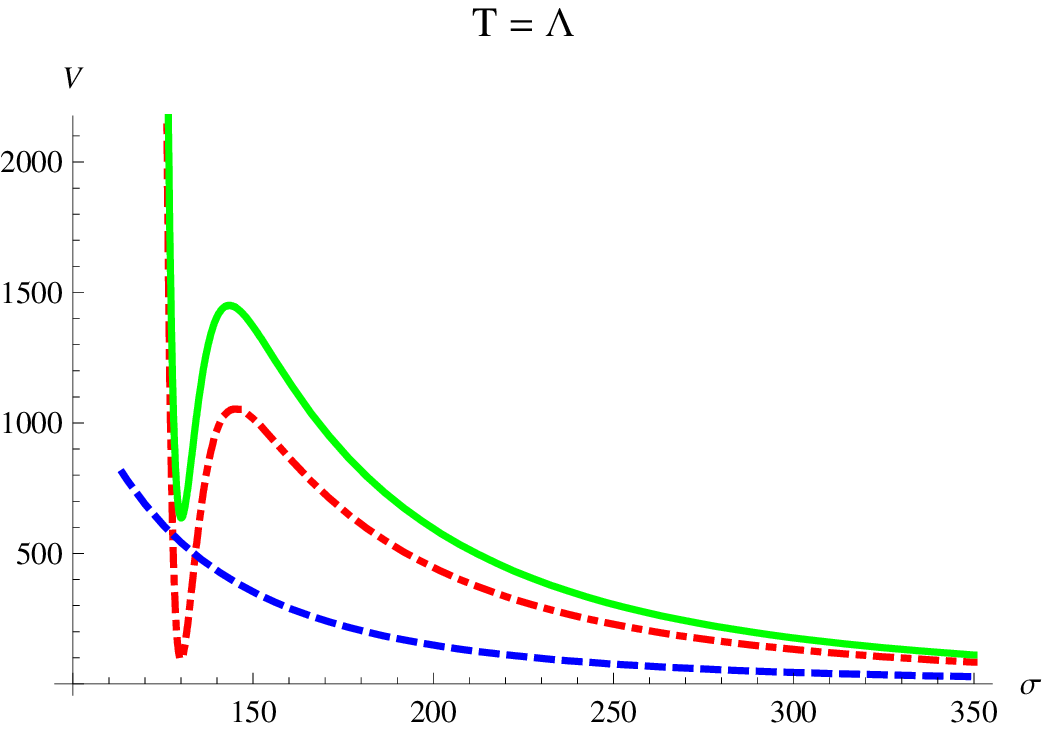}} \hspace*{0.3cm}\scalebox{0.72}{\includegraphics{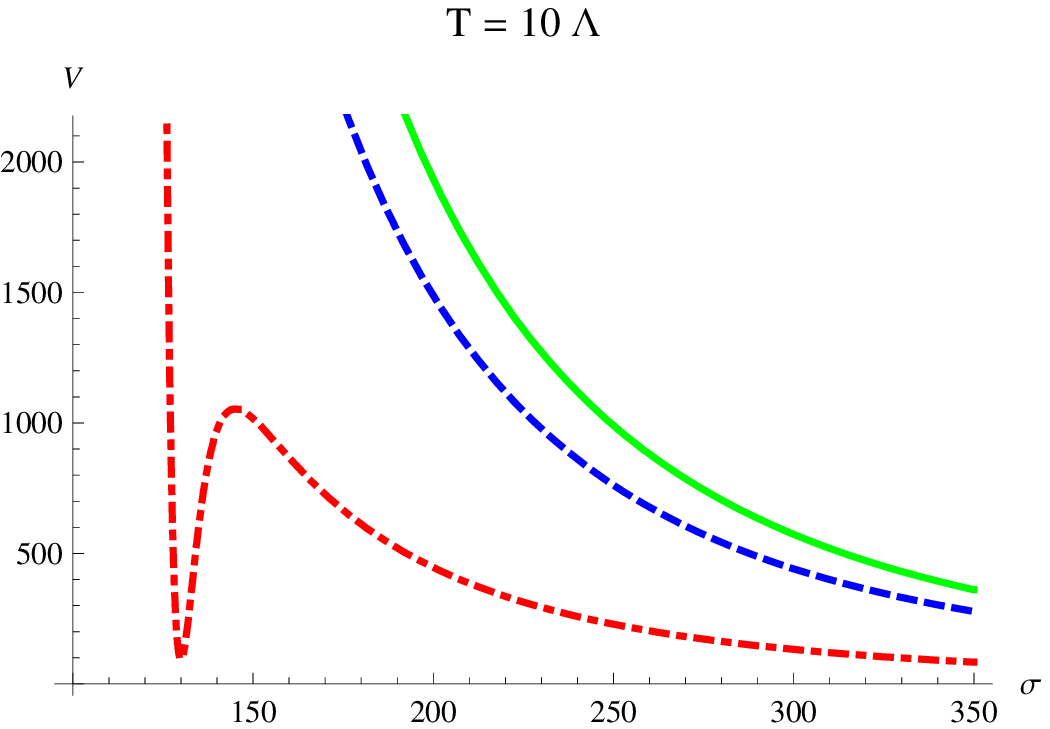}}
\end{center}
\vspace{-0.5cm}
\caption{{\small Effective potential (multiplied by $10^{30}$) for the one exponential case (green continuous line) compared to $V_0$ (red dot-dashed line) and $V_T$ (blue dashed line) for $T \approx \Lambda$ (left) and $T \approx 10 \Lambda$ (right). The values of the parameters are the following: $a=0.2$, $W_0=-10^{-10}$, $\mu = 1.3 \times 10^{-5}$, $\Lambda=10^{-5}$.}}
\label{TL}
\end{figure}

Now, for the largest allowed value of $\mu$ (i.e. $\mu\sim 10^{-2}$) the position
of the dS minimum is of ${\cal O} (10)$ and it increases as $\mu$ decreases, see Table \ref{table2}.
\begin{table}[h]
\begin{center}
\vspace{0.4cm}
\begin{tabular}{|c|c|c|}
\hline
$W_0$ & $\mu$  & $x_{dS}$ \\
\hline
$-10^{-4}$ & $1.3 \times 10^{-2}$ & 11.6 \\
\hline
$-10^{-6}$ & $1.3 \times 10^{-3}$ & 16.4 \\
\hline
$-10^{-8}$ & $1.3 \times 10^{-4}$ & 21.2 \\
\hline
$-10^{-10}$ & $1.3 \times 10^{-5}$ & 26.0 \\
\hline
\end{tabular}
\end{center}
\vspace{-0.2cm}
\caption{ {\small Each row in this table is a set of parameters for which $V_0$, in the case with one exponential, has a dS vacuum at $x_{dS}$.} \label{table2}}
\vspace{0.1cm}
\end{table}
Hence the two ratios of quadratic polynomials in (\ref{VTV0R}) are
of ${\cal O} (1)$. Therefore, the magnitude of
$V_T^{(0)}/V_0^{(0)}$ at $x_{dS}$ is determined by the magnitudes
of $T^2/\Lambda^2$ and $T^2 x_{dS}^2$. As $\Lambda<\!\!<1$ (i.e.
$\Lambda$ is at most of order $10^{-2}$), we always have that
$\Lambda^{-2}$ is at least of order $10^4$. Thus, when $x_{dS}$ is
of ${\cal O} (10)$, the first term in (\ref{VTV0R}) is dominating
and so the $V_0$ minimum persists in $V_{eff}$ as long as
$T<\Lambda$, whereas for $T>\Lambda$ the effective potential has
the runaway behaviour of $V_T$; see Figure \ref{TL} for an
example.\footnote{Recall that these inequalities are only
order-of-magnitude estimates; the transition does not have to
happen precisely at $T=\Lambda$, only at $T$ that is of ${\cal O}
(\Lambda)$.} For $x_{dS}$ of ${\cal O} (100)$ both terms
contribute with equal importance and for $x_{dS}$ of ${\cal O}
(10^3)$ or greater the condition for the existence of a finite-$x$
minimum of $V_{eff}$ is $T < x_{dS}^{-1}$.

Note that $x_{dS}$ is independent of $\Lambda$, but it depends on $\mu$ and $W_0$. As the
values of the latter two parameters are decreased the value of $x_{dS}$
increases, as can be seen in Table 2. One can verify that the position of the minimum is of ${\cal O}
(100)$, for example, for $W_0=-10^{-30}$ and $\mu=1.4 \times 10^{-15}$, in which case
$x_{dS} = 73$. However, such small parameters would require huge fine-tuning and so it is
preferable to consider the regime in which $W_0$ and $\mu$ are as large as possible. As we
saw above, in this regime $x_{dS}\sim {\cal O} (10)$ and so the condition for the
existence of a local minimum of $V_{eff}$ at finite $x$ is $T<\Lambda$. Then, clearly, for $\Lambda \sim {\cal O}(10^{-2})$ the finite-$x$ minimum of $V_0$ is not washed out by $V_T$ in the whole range of validity of supergravity. For any smaller $\Lambda$, though, this is not the case; the critical temperature above which $V_{eff}$ has a runaway behaviour is well within the range $T<\!\!<M_P$ and so one can reliably conclude that there is a phase transition towards the minimum at infinity.

\section{Conclusions}

We studied the one-loop temperature corrections to the effective
potential of the O'KKLT model. It turns out that, when the
non-perturbative superpotential contains only one exponent, the
temperature dependent part $V_T$ has runaway behaviour. For a
superpotential with two exponentials, on the other hand, $V_T$
{\it can} have minima at finite field vevs. Surprisingly however,
the minima of $V_{eff}$ still turned out to be determined by the
minima of the zero-temperature contribution, in the range of
validity of supergravity. So, although our initial motivation was
to see how the system evolves as $T$ decreases, assuming that it
started at a finite-vevs minimum of $V_T$, we ended up with a
quite different interpretation of our results. Namely, that there
is a regime in which reheating does not destabilize the
zero-temperature de Sitter minimum of the volume modulus $\rho$ in
the O'KKLT model (in the supergravity approximation). Note that
the existence of this regime is in stark contrast with the
situation in \cite{FKKMT} (and the coupling of that model to
supergravity \cite{ART}), where the relevant critical temperature
was found to be $T_c<\!\!<M_P$. The main differences between their
model and the one we studied here are that in our case the
superpotential contains exponentials, in addition to polynomials
in the fields, and the K\"{a}hler potential is non-canonical. It
would be interesting to understand which of these two features is
more essential, and to what degree, in order to obtain the kind of
result that we did.

It is of great interest to determine whether our conclusions about the finite temperature  behaviour of the O'KKLT model persist in more realistic cases, like the large-volume compactifications of \cite{Quevedo}. One would also like to understand corrections, that are due to taking into account of higher derivative terms appearing in effective supergravity actions. Another important problem is to get a handle on dynamical phenomena, in particular on the Jeans instability. A more fundamental question is related to the following issue: In \cite{BG1,BG2} the classical $T=0$ supergravity contribution to the $T\neq 0$ effective potential is itself viewed as an effective potential, in which the $T=0$ loop corrections are already taken into account. However, there are important subtleties in regulating the $T=0$ quadratic divergences in a supersymmetric manner (see the last five references in \cite{OneLoop}), which subtleties can have an impact on, for example, the phenomenology of flavor-changing neutral currents \cite{GN}. It is conceivable then, that also at $T\neq 0$ (despite susy being broken) the proper regularization may affect quantities of phenomenological interest. This is a rather important open problem and we hope that our unexpected result raises additional interest in it.

Finally, it would be very interesting to consider thermal corrections in a moduli stabilization set-up that requires going beyond the supergravity approximation, such as the non-geometric compactifications of \cite{BBVW}, so that one has to use string theory at finite temperature.

\section*{Acknowledgements}
We would like to thank M. Douglas, A. Iglesias, M. Ro\v{c}ek and S. Thomas for  discussions. We are also grateful to S. Thomas for reading the draft and providing useful comments.
L.A. thanks the Les Houches 2007 summer school "String Theory and the Real World" and the 5th Simons
workshop in Mathematics and Physics, Stony Brook 2007, for hospitality during the completion of this project. The work of
L.A. and V.C. is supported by the EC Marie Curie Research Training Network MRTN-CT-2004-512194
{\it Superstrings}.

\appendix

\section{Counting solutions of $e^{x} = F(x)$}
\setcounter{equation}{0}

In the main text, we will encounter on plenty of occasions equations of the type
\be \label{MainEq}
e^{cx} = F(x) \, ,
\ee
where $x$ is a real variable, $c$ is a constant and $F(x)$ is an expression containing (ratios) of polynomials and possibly other exponentials. In general, the analytic solution of this equation is not known.\footnote{In the very simple case $F=x^m$ it is. However, we will have to deal with significantly more complicated functions $F$.} Nevertheless, one can find an upper bound on the number of its solutions, as we explain below.

For simplicity, let us take $c=1$ in the rest of this appendix; the generalization for arbitrary $c$ will be obvious. Since $e^x$ is a monotonically increasing  function, if $F$ were monotonically decreasing (and continuous, which will always be the case) then clearly there could be only one or zero solutions depending on whether the value of $F$ is greater or smaller than the value of the exponential in the beginning of the interval of interest. The difficult case to analyze is when $F$ is also monotonically increasing; we will turn to it in a moment. Generically, in the cases of interest for us $F$ will not be monotonic. However, one can split the interval, that one wants to solve (\ref{MainEq}) in, into subintervals in which it is monotonic by considering the equation\footnote{Here we are assuming that $F(x)$ does not diverge anywhere inside the interval of interest. Otherwise an additional division into subintervals is necessary which, although complicating the considerations, does not lead to anything new conceptually.}
\be \label{EFd}
F'(x) = 0 \, ,
\ee
where $'$ denotes $d/dx$. Let us denote by $y_1, ..., y_m$ the solutions of (\ref{EFd}), where for convenience we have assumed the ordering $y_j < y_{j+1}$ for every $j=1,...,m$. In each interval $(y_j, y_{j+1})$ the function $F$ is monotonic: monotonically increasing if $F'>0$ for $x\in (y_j, y_{j+1})$ and monotonically decreasing if $F'<0$ for $x\in (y_j, y_{j+1})$.\footnote{In fact, in mathematics  the term 'monotonic' refers to functions for which $F' \ge 0$ or $F' \le 0$. The case when $F' > 0$ or $F' < 0$ is called 'strictly monotonic'. Since in our context it is clear what we mean, we will drop the adjective 'strictly' so as not to burden the language unnecessarily.} As already mentioned above, the intervals in which $F$ is decreasing are trivial to analyze. So let us from now on consider an interval $(y_k, y_{j+1})$ such that in it $F'>0$.

To recapitulate, we are considering now the equation
\be \label{NEq}
e^x = F(x)
\ee
in an interval $(y_k, y_{k+1})$ such that
\be
\forall x\in (y_k, y_{k+1}) \, : \qquad F'(x) > 0 \, .
\ee
In other words, in the interval of interest the function $F$ does not have any extrema or inflection points (nor any divergences except possibly at the end points $y_k$ and $y_{k+1}$) and is monotonically increasing. Let us denote the solutions of (\ref{NEq}) by $x_1,...,x_n$.\footnote{Unless stated otherwise, from now on we mean solutions in the interval $(y_k, y_{k+1})$.} Again we assume the ordering $x_i< x_{i+1}$ \,$\forall i=1,...,n$. We should also mention that we are considering only continuous functions, i.e. both $F$ and $F'$ are continuous.

\begin{figure}[t]
\begin{center}
\hspace{-0.3cm} \scalebox{0.8}{\includegraphics{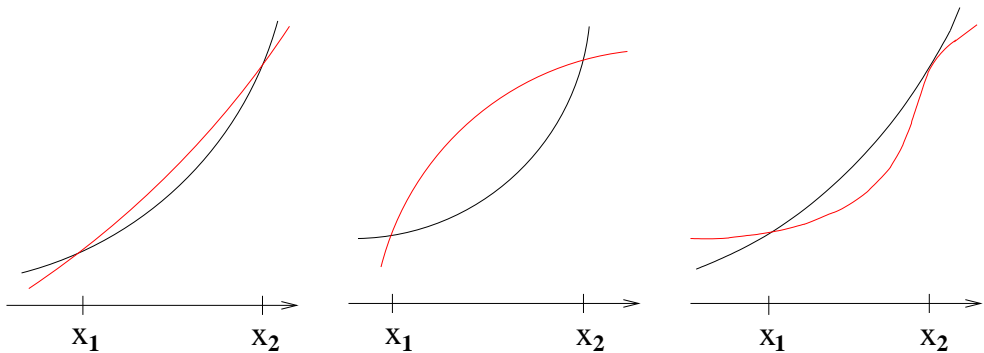}}
\end{center}
\vspace{-0.5cm}
\caption{\small Schematic depiction of the possible ways of intersection of two monotonically increasing functions, $f_1(x)$ and $f_2(x)$, at two successive solutions of $f_1(x)=f_2(x)$ for the case when $f'_1(x)=f'_2(x)$ has a single solution in the interval $(x_1,x_2)$.}
\label{Curves}
\end{figure}

Now, let us take two successive solutions, say $x_1$ and $x_2$. If at $x_1$ the derivative of one side of (\ref{NEq}) is greater than the derivative of the other, say
\be \label{AE1}
F'(x_1)>e^{x_1} \, ,
\ee
then clearly at $x_2$ the opposite inequality, or at least equality, has to be satisfied, i.e. $F'(x_2) < e^{x_2}$ or $F'(x_2) = e^{x_2}$. (Think of the tangents to the curves that represent the graphs of the functions at the two intersection points; see Figure \ref{Curves}.) Let us first consider the case when
\be \label{AE2}
F'(x_2) < e^{x_2} \, .
\ee
Since $F'$ is a continuous function, equations (\ref{AE1}) and (\ref{AE2}) imply that there has to exist a point $t\in (x_1, x_2)$ such that $F'(t) = e^{t}$. In other words, between two successive solutions of (\ref{NEq}) there is at least one solution of
\be \label{DerEq}
e^x = F'(x) \, .
\ee
\indent
Although slightly less obvious, the same conclusion can be reached also for the case when $F'(x_2) = e^{x_2}$. Indeed, if $F'(x) > e^x$ at $x_1$ and at every point between $x_1$ and $x_2$, then it is not possible that at $x_2$ the two functions are equal. (Think of two points moving on a straight line, the vertical axes, with different velocities. If they start from the same place at a moment of time $x_1$ and one is always slower than the other up until the moment $x_2$, then it is not possible that they meet at the moment $x_2$.) It has to be true that at least in some part of the interval $(x_1, x_2)$ the opposite inequality $F'(x) < e^x$ is satisfied in order for $x_2$ to be a solution. Therefore, again there has to be a point in between $x_1$ and $x_2$, in which the derivatives of the two functions are equal.

\begin{figure}[t]
\begin{center}
\hspace{-0.3cm} \scalebox{0.8}{\includegraphics{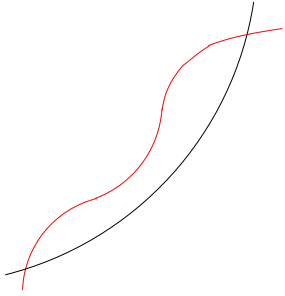}}
\end{center}
\vspace{-0.5cm}
\caption{ {\small Schematic depiction of two monotonically increasing functions $f_1(x)$ and $f_2(x)$, which exemplifies how the equation $f_1'(x) = f_2'(x)$ can have more than one solution between two successive solutions of $f_1(x) = f_2(x)$.}}
\label{Wobble}
\end{figure}

So we conclude that between each two successive solutions of
(\ref{NEq}) there has to be at least one solution of
(\ref{DerEq}). Let us denote the number of solutions of the latter
equation by $p$. Then the above considerations are summarized by
the following statement about the number of solutions, $n$, of
(\ref{NEq}): \be \label{UB} n \le p+1 \, . \ee One reason that the
RHS of (\ref{UB}) is just an upper bound and not the exact number
of solutions of (\ref{NEq}) is the possibility, that we already
considered above, for a solution $x_l$ of (\ref{NEq}) to also be a
solution of (\ref{DerEq}).\footnote{Clearly, if two successive
solutions of (\ref{NEq}) solve (\ref{DerEq}) too, then the
inequality in (\ref{UB}) only gets stronger.} Another is that the
function $F$ could 'wobble' as in Figure \ref{Wobble}
{}\footnote{Recall that in general its second (and higher)
derivative(s) is (are) also nontrivial function(s) of $x$.} and so
there could be more than one solution of (\ref{DerEq}) between two
successive solutions of (\ref{NEq}).

It is clear now what is the algorithm for counting (or rather
putting an upper bound on) the number of solutions of $e^x =
F(x)$. Namely, first find the solutions $\{y_j\}_{j=1}^m$ of
$F'(x) = 0$ and then in each interval $(y_j, y_{j+1})$, where
$y_0$ and $y_{m+1}$ are resp. the beginning and the end of the
interval in which we are solving $e^x=F(x)$, count the solutions
of $e^x = F'(x)$. If $F'$ is still a complicated function, it may
not be immediately obvious that it is a significant improvement to
consider the latter equation rather than the original one.
However, clearly one can develop a recursion, i.e. as a next step
view $e^x = F'(x)$ as the starting point for the above
considerations and so find the solutions of $F''(x) = 0$. Then in
each interval between two successive ones look for the number of
solutions of $e^x = F''(x)$ etc. until one reaches a rather simple
equation.\footnote{Of course, here we assume that all relevant
derivatives of $F$ are continuous.} This procedure is exactly the
tool that enables us in Appendix B to prove that there are no
local minima of $V_T$ for finite $\rho$ in the original KKLT
proposal (i.e., with a non-perturbative superpotential given by a
{\it single} exponential) for the range of parameters for which
the zero temperature potential has a dS minimum.

\section{No finite-$\rho$ minima of $V_T$ for one-exponential case}
\setcounter{equation}{0}

In this Appendix we show that the equation
\be \label{E1}
e^{2 x} = - C^{-1} \left( P_5(x) + W_0 P_3(x) e^x\right) \equiv H(x)
\ee
does not have any solution whenever $|W_0|$ and $\mu^2$ are of the same order of magnitude,
which is the condition (\ref{W0murel}) for the presence of dS vacua with a small
cosmological constant at zero temperature.

Following the logic of Appendix A, we consider as a first step the equation $H'(x) = 0$, where $'$ denotes differentiation w.r.t. $x$. The strategy is to split the half axes $[0, \infty)$ into intervals in which the function $H$ is monotonic (the intervals between successive solutions of $H'=0$ plus, of course, the interval to the left of the smallest solution and the interval to the right of the largest one) and find bounds on the number of solutions of (\ref{E1}) in each such interval.

Let us write the equation $H'=0$ in the form:
\be \label{Eq1}
e^x = - \frac{1}{W_0} \frac{29 + \frac{242}{3} x + 86 x^2 + 56 x^3 + 20 x^4}{38 + 67 x + 37 x^2 + 6 x^3} \equiv - \frac{1}{W_0} \frac{P_4}{\hat{P}_3} \equiv h(x) \, .
\ee
Clearly, we are again facing a transcendental equation of the type (\ref{MainEq}). Nevertheless, things have improved significantly since now we have on the RHS a function that we can fully analyze. As all coefficients of both polynomials $P_4$ and $\hat{P}_3$ are positive (and zero is obviously not a root of any of them), $h(x)$ does not diverge anywhere in the interval $[0, \infty)$ and is a monotonic function there. Hence (\ref{Eq1}) has a single solution when $h(x)|_{x=0}\ge e^x|_{x=0} = 1$.\footnote{For an arbitrary function $h(x)$ this would only be true if it were monotonically decreasing, unlike the case at hand. However, the rational function $P_4/\hat{P}_3$ approaches the behavior of its asymptote, $ \frac{10}{3} x$, around $x\sim 10$ and so if it does not cross $e^x$ for smaller $x$ it never does. One easily sees, by plotting the two functions for $x\in [0,\,10]$, that they  can intersect only if $h(x)$ was the greater function at $x=0$. \label{Pl}} So we find that $-0.76 \le W_0 < 0$. Since for the uplifting of the local zero-temperature minimum to dS one needs $|W_0| \approx \mu^2 <\!\!<1$, clearly for the case of interest for us the equation $H'(x)=0$ has one solution. Let us denote it by $x_0$. One can find $x_0$ numerically\footnote{As equation (\ref{Eq1}) depends on the value of $W_0$ so does $x_0$. For $|W_0| = 10^{-1}, 10^{-2},..., 10^{-6}$ it is  $x_0 = 4.6$, \,$7.5$, \,$10.2$, \,$12.7$, \,$15.3,$ and $17.7$ respectively. However, the conclusions of our subsequent analysis are the same for any $|W_0| < 0.76$. \label{FN}} and one can verify that $H'(x)<0$ for $x<x_0$ and $H'(x)>0$ for $x>x_0$.

Hence we have determined that there are two intervals in which the function $H(x)$ is monotonic: $[0, x_0)$ and $[x_0, \infty)$. In fact, we can discard the first of them immediately. The reason is that $H'<0$ in it and that in its beginning $H(0)<0$ for any $W_0$ in the range of interest (as mentioned above, this means $|W_0|<\!\!<1$). Therefore $H(x)$ remains negative throughout the whole interval and so cannot be equal to $e^{2x}$ at any $x$ there. So from now on we will only consider $x\in [x_0,\infty)$ .\footnote{Clearly, we could shift upwards the lower end of the interval of interest by taking it to be the point $t$ at which $H(t) = 0$, since obviously $t>x_0$. However, for our subsequent considerations it will not matter whether we consider $[x_0, \infty)$ or $[t, \infty)$. So we do not bother determining $t$.}

According to Appendix A, we have to aim now at counting the number of solutions of $(e^{2x})'=H'(x)$ in this interval. However, this is still a rather complicated equation. So we go to the next level of iteration by considering it as the starting point and looking for the solutions of $H''(x) = 0$ in order to find the intervals in which $H'$ is monotonic. For that purpose, let us write $H''(x) = 0$ in the following way:
\be \label{Eq2}
e^x = - \,\frac{1}{W_0} \,\frac{\frac{242}{9}+\frac{172}{3} x+56 x^2+\frac{80}{3} x^3}{35+47 x + \frac{55}{3} x^2 + 2 x^3} \equiv r(x) \, .
\ee
It is obvious that, similarly to $h(x)$, the function $r(x)$ does not diverge anywhere in the interval of interest and is monotonic in the whole of it. However, at the beginning of this interval $e^{x_0} > r(x_0)$ and hence (\ref{Eq2}) does not have any solution.\footnote{Similar remark as in footnote \ref{Pl} applies here.} Then one easily verifies that $H''(x)>0$ for any $x\in [x_0, \infty)$. Unfortunately though, the equation $(e^{2x})''=H''(x)$ is still not simple enough for us to be able to count its solutions. So we have to go to the next level, i.e. look for zeros of the third derivative, $H^{(3)} (x)$.

However, at this point it is clear how the iteration procedure will converge. Each further derivative decreases the power of the polynomial in the numerator, until what started as $P_5(x)$ completely disappears. At the level of $H^{(6)}(x) = 0$ one finds the equation $ 2^6 e^{2x} = e^x Q_3(x)$, where $Q_3(x)$ is still a degree-three polynomial. Clearly then this becomes $2^6 e^x = Q_3$ and after 3 more differentiations one finds $e^x = const$, which can have at most one solution. Also, at each step of the procedure the corresponding derivative of $H(x)$ is easily seen to be positive-definite in the whole interval of interest. Therefore, from Appendix A it follows that for the moment we have restricted the number of solutions of (\ref{E1}) to be at most 1+6+3=10.

At first sight this may not seem very encouraging. However, recall
that at the last step we arrived at an equation that can have at
most one solution. More precisely, it is \be \label{LE} 2^6 e^x =
\frac{12 |W_0|}{C}. \ee Now, since \be C = 3
\,\frac{\mu^4}{\Lambda^2} + \frac{3}{2} W_0^2 \, , \ee it is easy
to see that (\ref{LE}) does not have a solution neither for the
uplifted case (in which $|W_0| \approx \mu^2$ and so $C >\!\!>
W_0^2$) nor for the case with no uplifting (for example, take
$\mu=0$ and so $C\approx W_0^2$).\footnote{For instance, for
$W_0=-10^{-4}$ we have $x_0 = 12.7$ (see footnote {\ref{FN}}) and
so the LHS of (\ref{LE}) is $2^6 \times 3\times 10^5$, whereas the
RHS is $4\times 10^{-2}$ for the uplifted case with
$\Lambda=10^{-3}$ and $8\times 10^4$ for no uplifting with
$\mu=0$. In fact, one can easily convince oneself that our
considerations are independent of the particular value of $W_0$
(and the resulting value of $x_0$). Namely, one can check that
although the ratio between the LHS and RHS of (\ref{LE}) varies
for the six values of $W_0$ in footnote \ref{FN}, its order of
magnitude remains the same for all six values. (For example, for
the uplifted case one always finds that LHS/RHS $\sim$ $10^8$.)}
This in turn means that the equation $(e^{2x})^{(8)}=H^{(8)}(x)$
can have at most one solution. And since $H^{(8)}$ is monotonic in
the whole interval of interest, it is again very easy to verify
that it too does not have any. Therefore
$(e^{2x})^{(7)}=H^{(7)}(x)$ can have at most one solution etc..
Actually, we should mention that the equation
$(e^{2x})^{(7)}=H^{(7)}(x)$ is the stage at which a difference
appears between the cases with and without uplifting. Namely, for
the case with no uplifting the two sides of this equation are of
the same order of magnitude at the point $x_0$; for the lower
derivative equations the side of the exponential is the smaller
one and hence we cannot decrease the bound on the number of
solutions of $e^{2x} = H(x)$ any further. On the other hand, for
the uplifted case we find that there are no solutions at each step
until $(e^{2x})' = H'(x)$ and finally $e^{2x} = H(x)$ itself.

\end{document}